# Experimental realization of Majorana hinge and corner modes in intrinsic organic topological superconductor without magnetic field at room temperature

Kyoung Hwan Choi[1], and Dong Hack Suh[1†]

1 Advanced Materials & Chemical Engineering Building 311, 222 Wangsimni-ro, Seongdong-Gu, Seoul, Korea, E-mail: dhsuh@hanyang.ac.kr

**Abstract**

Exotic states of topological materials are challenging or impossible to create under ambient conditions.[1-4] Moreover, it is unclear whether topological superconductivity, as a critical element for topological quantum computing, exists in any naturally occurring materials.[5-7] Although these problems can be overcome through the combination of materials in heterostructures, there are still many requisites, such as low temperatures and specific magnetic fields.[8,9] Herein, an intrinsic topological superconductor that does not depend on particular external conditions is demonstrated. It is accomplished utilizing the unique properties of polyaromatic hydrocarbons (PAHs), which have been proposed to have persistent ring current.[10-12] According to the Su-Schrieffer-Heeger(SSH)[13] and Kitaev[14] models, PAHs can have a non-trivial edge mode, so that perpendicularly stacked PAHs are expected to have Majorana hinge and corner modes.[15] Intrinsic persistent ring current of HYLION-12 is demonstrated by MPMS.[16] Coherent Quantum Phase Slip(CQPS), the Constant Conductance Plateau (CCP) and the zero bias conductance peak(ZBP) which is signatures of hinge modes are confirmed through the Josephson junction device of pelletized orthorhombic phase organic crystals of HYLION-12 by transport spectroscopy.[17,18] They are signatures of Majorana hinge and corner modes. In addition, the braidinglike operation by transport spectroscopy shows the emergence of the most important and critical elements of quantum computers that can be realized without an external magnetic field at room temperature.

**Introduction**

The discovery of the quantum spin Hall (QSH) effect has boosted the study of topological materials.[19-22] The QSH insulator, also known as a time-reversal symmetric topological insulator and characterized by a $\mathbb{Z}_2$ invariant[23,24], hosts a helical electron mode localized on the edges of the insulator[25]. This helical mode is stable under time-reversal symmetry and leads to the quantized two-terminal longitudinal charge conductance $G = 2e^2/h$, which is robust against disorder scattering and other perturbation effects.[26-28] The concept of robust boundary modes protected by symmetries stimulates the development of topological superconductors (TSCs) which are essential components of quantum computers.

There has been enormous interdisciplinary interest in TSCs. The first generation of TSCs consisted of a 1D superconductor (SC) hosting a Majorana bound state on the wire ends.[2,29,30] A Majorana bound state has zero energy bounded by particle-hole symmetry. The observation of the ZBP has provided tentative evidence supporting the existence of the Majorana bound state.[5,17,31-34] Recently, another important breakthrough showed evidence of the second-generation TSCs, as 2D TSCs hosting a chiral Majorana edge mode, through observation of the plateau of 0.5 $e^2/h$ and two terminal conductance in a hybrid quantum anomalous Hall-superconductor device.[35-38].

According to the classification table of topological insulators and superconductors,[39-41] the third-generation TSCs can be realized as 2D time-reversal symmetric TSCs hosting 1D Helical Majorana Edge Modes (HMEMs) in DIII class. An HMEM, composed of two chiral Majorana edge modes with opposite propagating directions as a Kramers pair, cannot hybridize in the presence of time reversal symmetry. The possible realization of HMEMs has been proposed in unconventional superconductors with exotic superconducting pairings[42-46] and the interface of two s-wave superconductors with a π-phase difference[47,48], and an HMEM is present on the edge of the s-wave superconducting proximitized antiferromagnetic quantum spin Hall insulator (AFQSHI) as an antiferromagnetic topological superconductor[49]. HMEM observation has not been realized, due to the subtlety and uncertainty of the superconducting pairing using unconventional superconductors and the difficulty in controlling the π phase without breaking time-reversal symmetry.[50,51]

Recently, higher-order topological insulators (HOTIs) and superconductors (HOTSCs) have attracted considerable interest owing to the emergence of unconventional bulk-boundary correspondence.[52-55] As is known, the boundary modes of conventional topological insulators (TIs) and TSCs are located at their one dimensional lower boundaries[4,56]; however, for a nth order TI or TSC with n≥2, the boundary modes are located at the n-dimensional lower boundaries. In the cases of 2D and 3D structures, such boundary modes are commonly dubbed corner or hinge modes. These modes have been predicated to exist in quite a few materials[57-61] and observed in various platforms, including bismuth[62] and iron-based superconductors[63]. A helical Majorana hinge mode is also proposed in 3D intrinsic higher-order topological odd parity superconductors (TOPSCs) based on second-order TOPSCs.[15]

On the other hand, the large anisotropic diamagnetic susceptibility exhibited by aromatic molecules, and the associated NMR frequency shifts at nearby atoms are generally interpreted to be arising from delocalized ring current in the molecular rings.[64] With respect to ring current, F. London suggested that the diamagnetic current in aromatic rings is analogous to supercurrent in superconductors.[10] After the advent of BCS theory, the possible link between ring current in aromatic molecules and persistent current in superconductors has been discussed by W. A. Little utilizing a proposed excitonic pairing mechanism.[11] Recently, J.E. Hirsch proposed that the large diamagnetic response of aromatic molecules may originate in the existence of a spin current in the ground state of such molecules in the absence of applied fields.[12] However, these hypotheses that aromatic compounds are superconductors have not been reported so far.

Here, we report the experimental results of helical Majorana hinge and corner modes in organic crystals. Their conductance shows a CCP and a ZBP, which simultaneously depicts the Majorana hinge and the non-local Majorana zero modes of odd parity. This result proves that the aromatic molecules constituting the organic crystals are both 2D intrinsic first and second-order TOPSCs and that the perpendicularly stacked crystals are 3D intrinsic higher-order TOPSCs. It is the first report that topological superconductors were implemented without an external magnetic field at room temperature.

**Results and discussion**

In recent years, researches on topological materials based on organic substances have been hardly conducted, whereas those on topological materials with inorganic ingredients have been actively conducted. However, the exotic conductivity of polyacetylene shows the importance of organic materials with regard to their topological aspects, and was investigated in more detail by the SSH model.[13,65] The model clearly explains that trivial and non-trivial states are generated in polyacetylene depending on the difference in the hopping amplitude of the intra- and inter-cell, and the concept is commonly adjusted to fit the Kitaev model[14]. Therefore, it can be expanded to polyaromatic hydrocarbons (PAHs), which can form non-trivial and trivial phases depending on the hopping amplitude because all carbons in PAHs also have three $sp^2$ hybrid orbitals for σ bonding and only one free $p_z$ orbital, as in polyacetylene. Furthermore, every benzenoid unit with three π orbitals can show a trivial state. (Fig. 1a) Otherwise, if the π orbitals are aligned in one direction (Fig. 1b), an edge state should be formed. Furthermore, magnon, a quantized spin wave, might be exist in HYLION-12. (Fig. 1c) This hypothesis is quite reasonable as this report shows that the ring current of PAHs is persistent and is accompanied by a spin flip.[12] Therefore, HYLION-12 has two different non-trivial states: one can have helical or chiral modes as a first-order topological material, and the other can have localized the quadrupole corner mode as seen in a higher-order topological material. Therefore, stacked HYLION-12 with quantized edges or corner modes is likely to be a 3D intrinsic

TOPSC.[15] (Fig. 1d)

Recently, the orthorhombic phase ($\Phi_o$) from HYLION-12, a pyrene derivative, has been reported to be a Dirac metamaterial.[66] It successfully demonstrated various topological photonic properties from the molecular topological superconductor. In addition, alignment to the z-axis of HYLION-12 in $\Phi_o$ is verified to 3D intrinsic higher-order topological odd parity superconductors (TOPSCs).[15] In this case, the quantized helical-edge conductance should be experimentally observed in mesoscopic samples.[67] If a non-trivial state exists in HYLION-12, it should be confirmed even in the mesoscopic state due to the robustness of the hinge mode. And if the persistent spin wave exists in the Majorana edge mode of HYLION-12, $\Phi_o$ from quantization between perpendicularly stacked HYLIONs-12 should have a helical hinge mode that has very large diamagnetic susceptibility in the absence of an external magnetic field. To confirm it, $\Phi_o$ is molded into a cylindrical shape with a diameter of 13 mm by pressing with 1, 1.5 and 2 metric tons (Fig. S1). Its crystal structure was characterized by XRD analysis. (Fig. 1e and 1f) All experiments were performed with the samples of $\Phi_o$ made at 1.5 metric ton pressure.

**Magnetic properties of $\Phi_o$**

The magnetic susceptibility of zero-field-cooled ($\chi_{ZFC}$) and field-cooled ($\chi_{FC}$) modes was measured at a constant magnetic field of 30 Oe. (Fig. 2a) The magnetic zero-field cooled curves (ZFC) were defined for samples that are first cooled (at H=0) to designated temperature before the external magnetic field is applied, whereas the magnetic field–cooled (FC) branches measure samples cooled to low temperature at 30 Oe. In both $\chi_{ZFC,30}$ and $\chi_{FC,30}$ cases, diamagnetic susceptibility is represented over entire temperature range, and the results reveal three exotic phenomena. One, in both cases, is that diamagnetic susceptibility increases with increasing temperature, whereas it is usually constant or gradually weakens as temperature rises.[68] It shows that HYLION-12's intrinsic ring current is nonresponsive to thermal perturbation. Moreover, the increase of ring current is due to the increase in temperature. Another peculiarity is that $\chi_{ZFC,30}$ is greater than $\chi_{FC,30}$. The ZFC curve generally lies below the FC curve up to the typical characteristic temperature representing a critical transition point corresponding to various physical states.[69] This data can also be understood in relation to HYLION-12's stable ring current. The external magnetic field applied at high temperature means that it is applied to electrons rotating at a higher velocity at high temperature than electrons already in circular motion at low temperature. According to Lenz's law, current is induced by the applied magnetic field, and electron motion increases faster. Therefore, $\chi_{FC,30}$ should have a larger diamagnetic susceptibility than $\chi_{ZFC,30}$. However, to keep induced current constant, ring current of HYLION-12 must be alternating current without resistance, even if the external magnetic field remains constant. Therefore, this make it possible to think of ring current as a quantized spin wave. (Fig. 1c) The other is that HYLION-12 exhibits a unique second order magnetic phase transition at 240K. This phenomenon appears to be deeply related to ring current of the PAH. Molecular local current of benzene is inherently favored by NNN hopping under relatively lower energy(lower temperature), and changed to NN hopping with continuity as temperature gradually increases.[70] Simultaneously the diamagnetic angular momentum due to ring current becomes larger as the orbit size decreases. Therefore, it can be expected that the path of ring current formed at temperature below 240K is related to NNN hopping of the HYLION-12, and that formed above 240K is related to NN hopping. The experiment confirms that ring current exists as alternating current in HYLION-12 and is not affected by thermal perturbation. Moreover, ring current increases as temperature increases, resulting in larger diamagnetic susceptibility.

Figure 2b shows the magnetic field dependence on a plot of M/H versus T. It is evident that the diamagnetic signal is dissipated by the external magnetic field. The results obtained at 0 Oe indicate the intrinsic magnetic characteristics of the orthorhombic phase. $\chi_{ZFC,0}$ shows diamagnetic susceptibility values of -11 x $10^{-3}$ emu $g^{-1}$ at 10K and -14 x $10^{-3}$ emu $g^{-1}$ at 300K. The value at 300K is 27.5 times greater than that of pyrolytic carbon which is generally considered to have very strong diamagnetic susceptibility (66.3 times greater than bismuth). However, when a weak magnetic field (30 Oe) is applied, $\chi_{ZFC,0}$ sharply decreases to -1.4x$10^{-5}$ emu $g^{-1}$. Despite this rapid decline, this value is similar to that of graphite.[71]

The experimental results significantly increase the likelihood that HYLION-12 and $\Phi_o$ are both superconductors with intrinsic ring current. Superconductors actively and perfectly isolates its interior from external magnetic fields, resulting in zero electrical resistance. Zero resistance implies that if one tried to magnetize superconductors, according to the Len's law, a current loop is generated to exactly cancel the imposed field, namely, there is no intrinsic current in superconductors. On the other hand, HYLION-12 may have its own persistent ring current and can support induced current with low resistance that is not affected by thermal perturbation.

Another interesting result is that the diamagnetic to antiferromagnetic transition is observed at 90 Oe. In the magnetic field range from 0 Oe to 70 Oe, diamagnetic susceptibility gradually decreases. However, above 90 Oe, antiferromagnetic susceptibility gradually increases. Diamagnetism originates from current formed by particles with a coherent wave function. In contrast, antiferromagnetism is derived from alignment in a regular pattern, with neighboring spins of different sublattices pointing in opposite directions.[72] Hence, the transition from diamagnetic to antiferromagnetic indicates that ring current of HYLION-12 slows down as the external magnetic field becomes stronger, and eventually disappears. Even without ring current, the spins are not aligned in one direction by the external magnetic field. Also, in general, antiferromagnetic material transition to paramagnetic as temperature increases above Neel temperature. However, this magnetic phase transition cannot be particularly observed even up to 320K. It indicates that the spins of HYLION-12 itself have aligned in a regular pattern with neighboring spins on different sub-lattices pointing in opposite directions, but it is not a result caused by temperature.

To further demonstrate the magnetic properties of $\Phi_o$ more deeply, the MH curve is measured by first increasing H from 0 Oe to 200 Oe. And a cycle of decreasing to -200 Oe and then increasing to 200 Oe is repeated three times at 200K. The MH plot clearly represents an exotic hysteresis curve. (Fig 2c) When the external magnetic field increases from -200 Oe to -60 Oe, the typical curve with antiferromagnetic susceptibility is displayed. (region I) However, in the region between -60 Oe and 60 Oe, negative constant magnetic susceptibility is constant irrespective of the change to the external magnetic field. This would be attributed to intrinsic ring current. (region II) When 60 Oe is exceeded, the changes in curves once again exhibit antiferromagnetic susceptibility. (region III) The same results are also observed when the external magnetic field decreases. Curves represent antiferromagnetic and diamagnetic susceptibility in regions IV and VI, and region V, respectively.

Experimental results acquired through MPMS again verify the existence of an intrinsic persistent ring current in HYLION-12. However, two things are not clear about what causes it. One is whether it comes from a quantized spin wave of HYLION-12 or not. The other is whether it comes from the periodic and regular structure of $\Phi_o$ and its overall quantization or not. Current and differential conductance can help to carefully characterize the property of $\Phi_o$.

**Transport spectroscopy with two terminals through the Josephson junction**

Current and differential conductance are measured by transport spectroscopy. The first column in Figure 3 shows the measurement set-up for the two-terminal experiment. The pelletized $\Phi_o$ has left and right contacts with normal leads at the top edge (red circle) or the side surface (cyan line). The surprising result is that constant current is present around zero bias, which is a fundamental characteristic of coherent quantized phase slip (CQPS) found at the Josephson junction of superconductors (Fig. 3II).[73] In addition, $I_1$ and $I_2$ have the same magnitude with opposite signs because when a potential bias applied to $V_1$, the hole is reflected as an electron at the interface between $\Phi_o$ and the normal lead, known as Andreev reflection (AR). Therefore, only electron current is measured at $I_1$, whereas only the hole current at $I_2$, and it happens from Cross Andreev reflection (CAR) (Supplementary information, cartoon). Another peculiar result is that when current through lead 2 is attached to the side surface, only exactly half of current is observed at the top edge. (Fig. 3c and d) This feature strongly suggests that current through $\Phi_o$ is caused from the hinge mode, since higher-order topological superconductors have different conductance depending on the terminal positions.

Differential conductance is more clearly identified in the non-trivial state of $\Phi_o$. What can be confirmed is the presence of a quantized conductance plateau between -257mV and 257mV in the top edge configuration. It is a signature of a Majorana hinge mode[17,49,74] and is the first band gap ($\Delta_{1,\text{top edge}}$). Besides, there are two conductance peaks with an odd function observed at -10mV and 12mV, and -2mV and 6mV. The result indicates a non-local Majorana corner mode. The non-local conductance is an odd function of bias voltage near the gap closing-reopening topological phase transition point.[75] This current rectifying effect is due to CAR.

Moreover, it can be observed that differential conductance oscillates by enlarging the constant plateau. (Fig. 3III inset) Analysis of MPMS experiments suggests the existence of non-dissipating spin waves in HYLION-12. It is experimentally ascertained through conductance oscillations.

It means $\Phi_o$ has both Majorana hinge and corner modes, and their concurrence is only possible when the Majorana corner mode of HYLION-12 connects continuously between the perpendicularly stacked layers in $\Phi_o$. In addition, current in $\Phi_o$ is observed at zero bias, showing that intrinsic persistent ring current of HYLION-12 is quantized throughout $\Phi_o$. Hence, HYLION-12 and $\Phi_o$ are two- and three-dimensional topological superconductors, respectively. And when the two-dimensional higher order topological superconductor is formed, the quadrupole should be identified. Thus, the quadrupole corner mode in HYLION-12 can be thought of as being formed by particles and antiparticles of up and down spins. According to this supposition, intrinsic current of HYLION-12 can be considered as the paired quantized spin waves, even in the absence of an external magnetic field. In addition, to have alternating current, a persistent spin texture is required, and it is also considered to be the corner mode in the spin direction that coincides with the plane containing the HYLION-12.

**Braidinglike operation of pelletized $\Phi_o$**

To understand the topological properties of $\Phi_o$ in more detail, current and conductance are measured by connecting four terminals. (Fig. 4) This experiment is conducted using two configurations: Van der Pauw, which is a method of measuring Hall conductance of mesoscopic materials without uniform in shape, and all terminals are connected directly to the ground.

At the top edge, when measuring using the van der Pauw method, $dI_3/dV_1$ is found to be the same as $dI_2/dV_1$ and $dI_4/dV_1$. (Fig. 4a III) It means that Hall current is rotating through the top edge of circular column samples. The measured $dI_3/dV_1$ is -0.14, which is 1/3 of the $dI_3/dV_1$ in the two-terminal experiment, and the total sum of conductance of all leads in the four-terminal experiment is the same as the $dI_3/dV_1$ in the two-terminal experiment. It is in accordance to the result of a braidinglike operation.[76]

When a quantum dot relates to electrical transport, two ribbons of hybrid TSC/QAHE junctions with opposite out of plane magnetization, the transport coefficients and terminal conductance can be affected by the potential of the QD, designated as $\varepsilon_d$ (See supplementary information). As $\varepsilon_d$ approaches zero, the transmission between two opposite chiral modes ($\gamma_1$ and $\gamma_2$) in different ribbons takes place with resonant exchanges of $\gamma_1 \to -\gamma_2$ and $\gamma_2 \to -\gamma_1$. As a result, the normal tunneling and AR coefficients from lead 1 to both leads 3 and 4 are equal. In other words, 2/3 of the hinge mode from lead 1 to lead 3 is separated, and it forms a circuit such that $\Phi_o$ connects leads 2 and 4 to create conductance. This effect shows the braidinglike operation in $\Phi_o$. Such a phenomenon is more clearly understood when all electrodes are directly connected to the ground. In this case, the same results are obtained as in the van der Pauw method. It means that conductance is precisely allocated even without a ground connection.

In contrast, when the van der Pauw experiment is performed on the side surface, there is no Hall conductance in leads 2 and 4. (Fig. 4c) It can be explained that there is no Hall current on the side surface. However, leads 1 and 3 have same current and conductance values, as in the two-terminal experiment. Also it indicates that CAR can occur at the hinge mode on the side surface, but Hall conductance does not exist on it. When all leads on the side surface are directly connected with ground, it shows that there are the same bridinglike operations as the leads connected to the top edge. (Fig. 4d)

From this, it can be surmised that Hall conductance on the side surface exists only in the vertical direction to connect top and bottom edges. To confirm this, two and four leads are attached on the side surface in a row, respectively. (Fig. 5a and b) Through the two-terminal experiment, CAR of the hinge mode is confirmed. And also, in the four-terminal experiment, the same values of current and conductance are measured in leads 2, 3 and 4. Simply put, it is verified that intrinsic current flows in the vertical direction of the side surface connecting top and bottom edges.

**Conclusion**

We found amazing phenomena in HYLION-12 and its crystal. First, aromatic compounds have intrinsic paring quantized spin waves. Its characteristic is ascertained from the exotic magnetic properties of $\Phi_o$ and HYLION-12 has large diamagnetic susceptibility even in the absence of an external magnetic field, and the MH curve clearly depicts both the induced and intrinsic persistent ring current of it. It is also characterized that intrinsic persistent current is truly quantized via transmission spectroscopy. The current-voltage curves exhibit the CQPS, and a CCP and a ZBP originating from the non-local Majorana corner mode are verified through differential conductance, it indicates that $\Phi_o$ is a 3D intrinsic higher-order TOPSC. It is possible because pyrene itself of the aromatic core of HYLION-12 is a 2D second-order TOPSC. What is even more surprising is that all of these experiments were carried out without an external magnetic field at room temperature. The discovery of a new quantum material, HYLION-12, will pave the way for a new era of quantum computers.

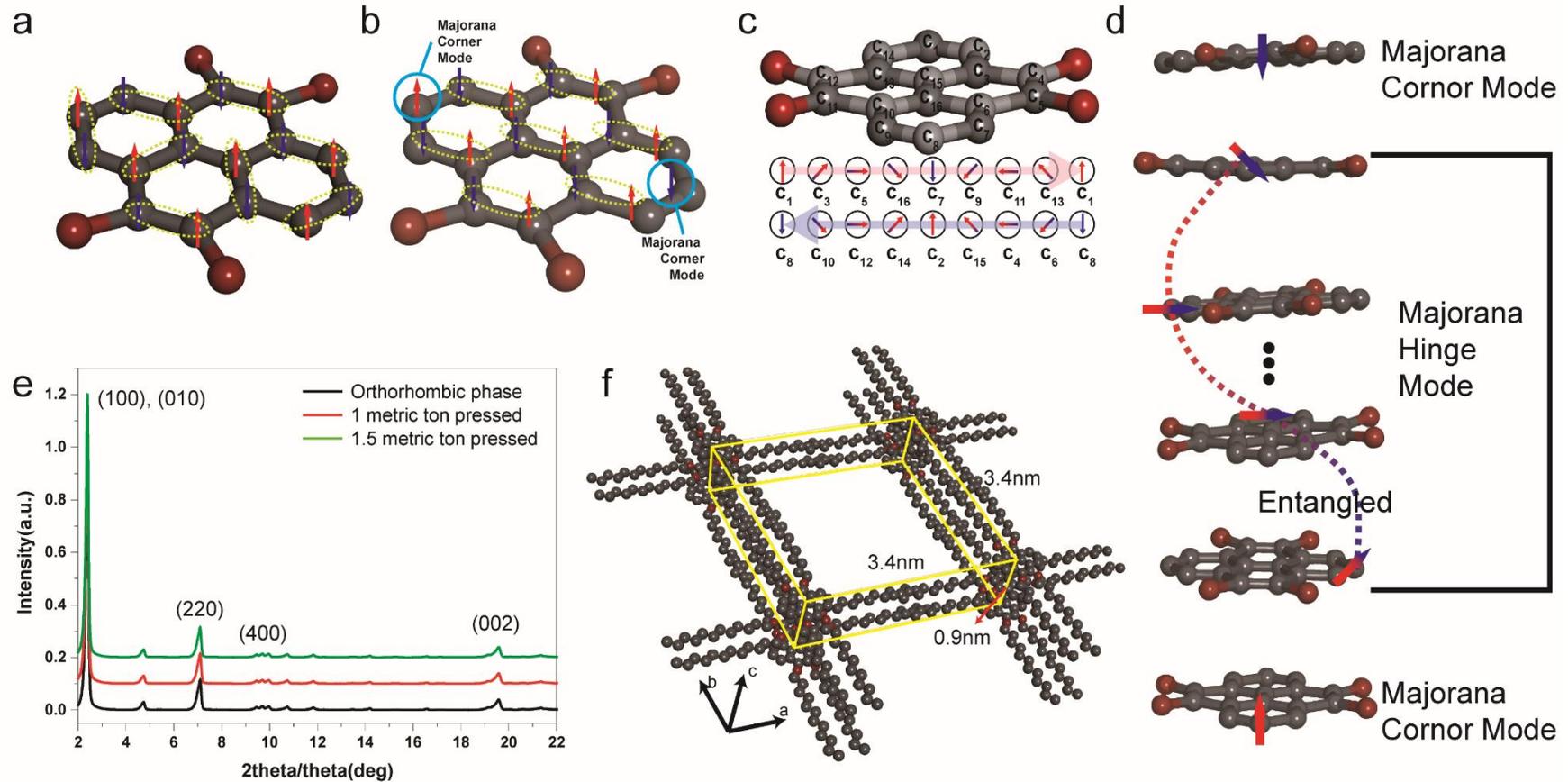

**Figure 1. Scheme of edge state of HYLION-12. a.** Trivial state of HYLION-12. **b.** Majorana corner mode as second order topological superconductor in HYLION-12 **c.** Quantized spin wave of HYLION-12 as first order topological superconductor. **d.** Majorana hinge and corner modes in stacked HYLION-12 as three dimensional intrinsic odd parity higher-order topological superconductor. **e and f.** XRD patterns of pelletized $\Phi_o$(e) and its structure by simulation(f).

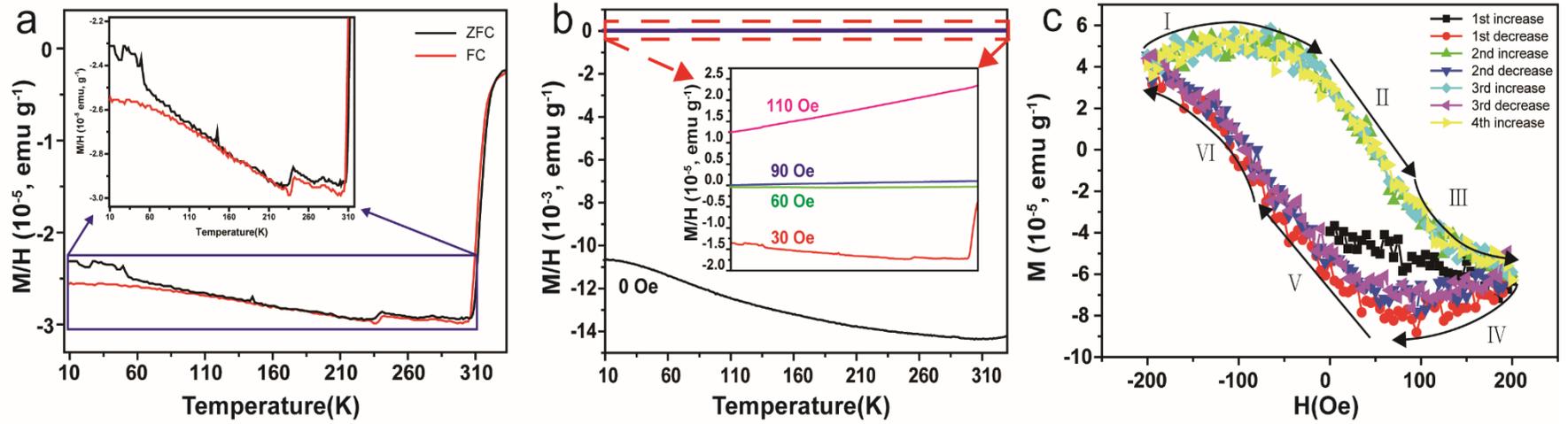

Figure 2. Magnetic property of pelletized $\Phi_o$. **a.** ZFC and FC curves in 30 Oe. **b.** ZFC curves with 0 Oe(black), 30 Oe(red), 60 Oe(green), 90 Oe(blue) and 110 Oe(magenta). **c.** MH curves of pelletized $\Phi_o$: first increase(black), first decrease(red), second increase(green) second decrease(blue), third increase(cyan), third decrease(purple) and fourth increase(yellow).

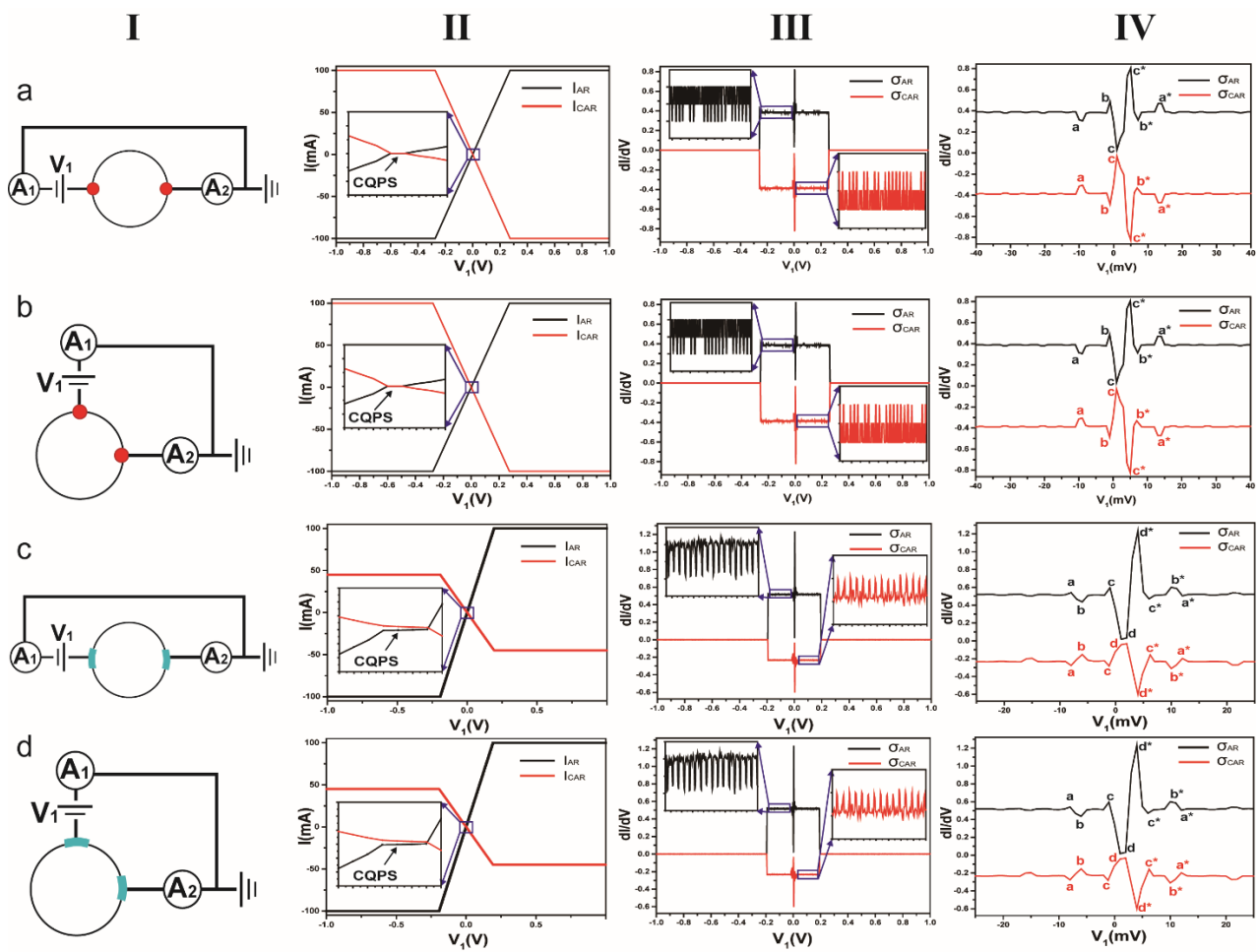

**Figure 3. Two terminal transport spectroscopy.** The first column(I) is configurations of experiment. Second column(II) is current-voltage curves. Third(III) and fourth(VI) column is differential conductance for CCP and ZBP. **Column II.** Coherent quantum phase slip (CQPS), phenomenon exactly dual to Josephson effect. It indicates that it coherently transfers vortices across superconducting wire. **Column III.** CCP and ZBP, signature of Majorana hinge and corner modes. Oscillation of conductance plateau proves quantized spin wave in HYLION-12. **Column VI.** Detail analysis of ZBP shows non-local relation of Majorana corner mode.

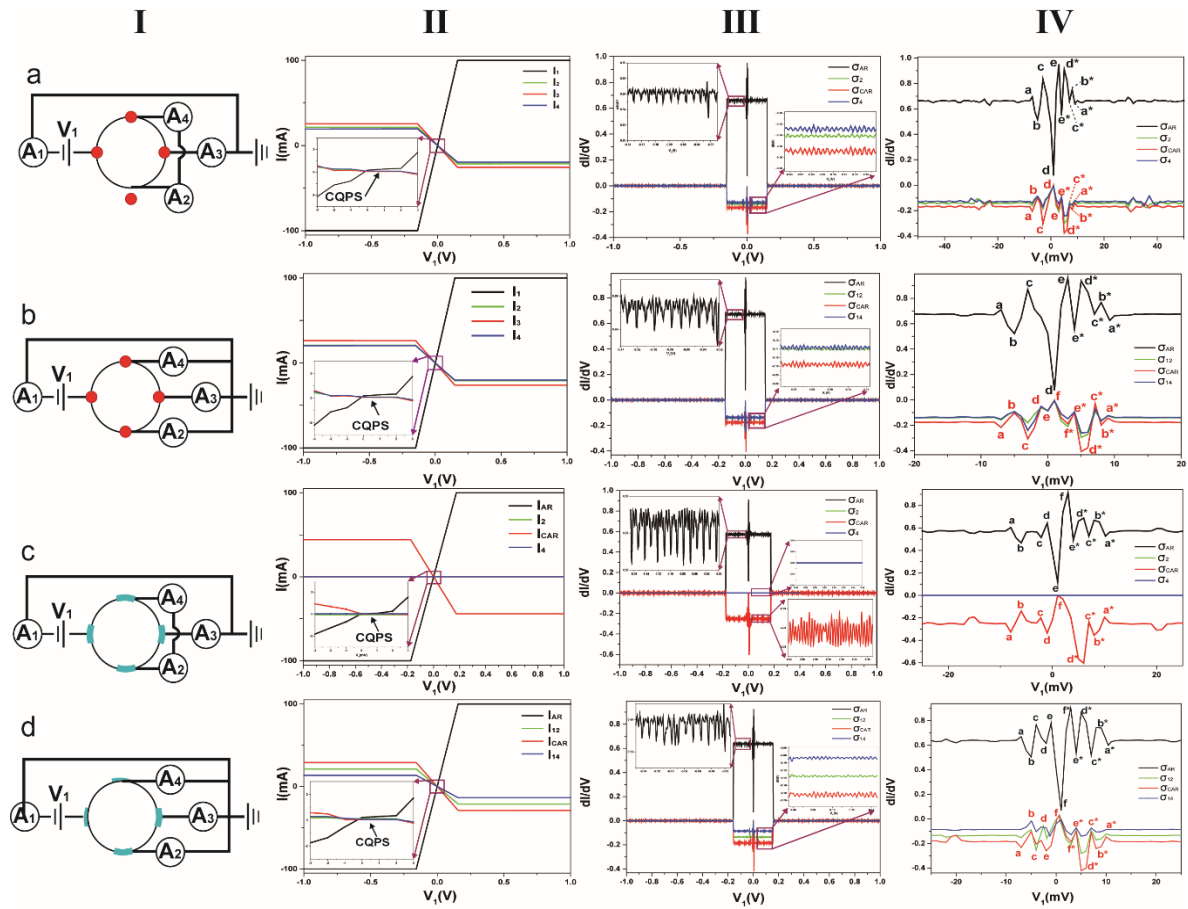

**Figure 4. Van der Pauw and four terminal transport spectroscopy.** The first column(I) is configurations of experiment. Second column(II) is current-voltage curves. Third(III) and fourth(VI) column is differential conductance for CCP and ZBP. **Column II.** There is no Hall current through side surface. (configuration c) **Column III.** Hall conductance is only confirmed on top edge configuration. Braidinglike operation of pelletized $\Phi_o$ is verified in whole configurations. **Column VI.** Local Majorana edge mode is only observed in configuration c: side surface has Majorana Corner Mode disconnected through Majorana hinge mode.

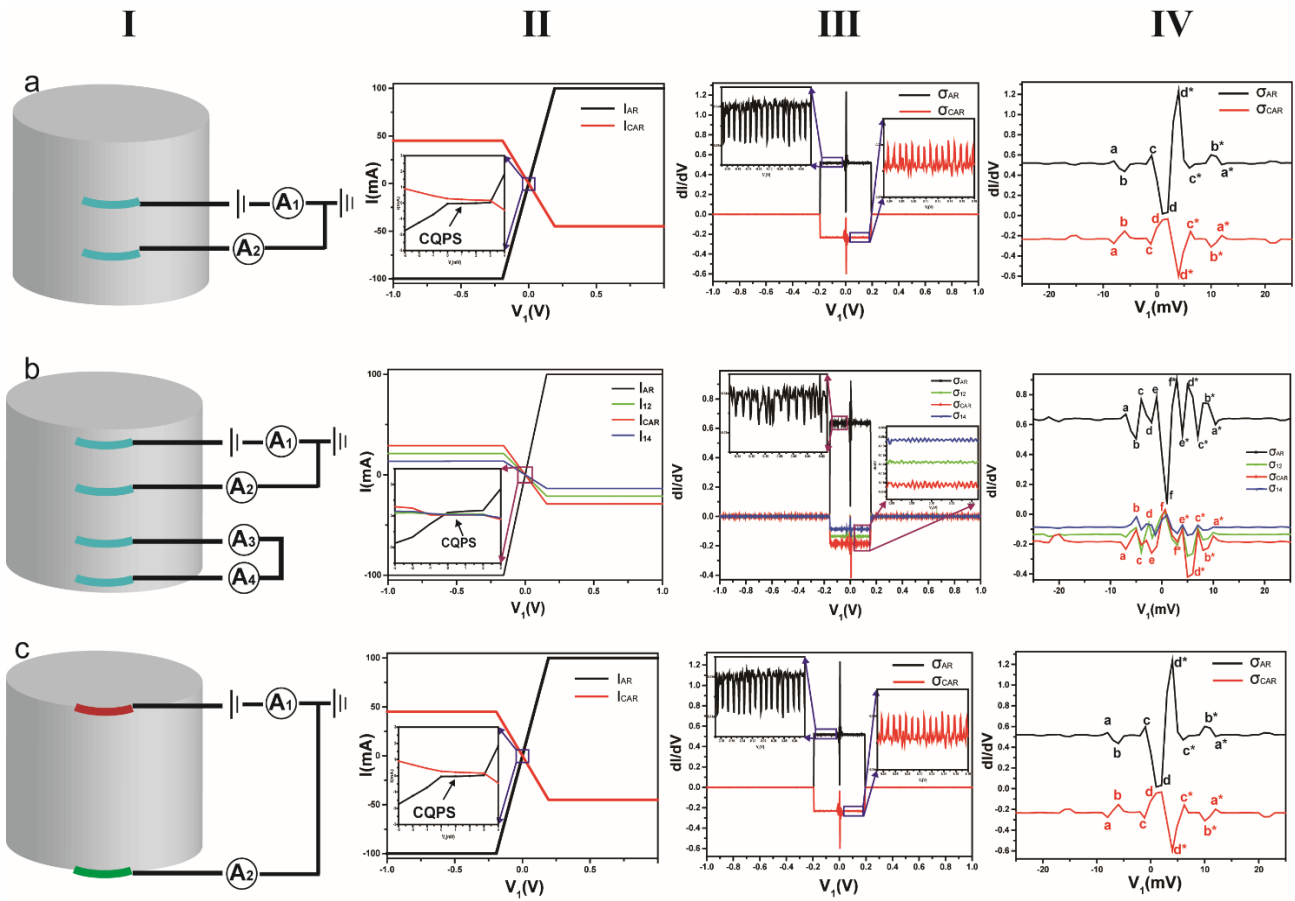

**Figure 5. Van der Pauw and four terminal transport spectroscopy for side surface.** The first column(I) is configurations of experiment. Second column(II) is current-voltage curves. Third(III) and fourth(VI) column is differential conductance for CCP and ZBP. **Column II.** It verifies that only half of current on top edge flows through side surface. Intrinsic current is confirmed on side surface. (configuration b) **Column III.** CCP and ZBP, signature of Majorana hinge and corner modes, are verified on side surface. **Column VI.** Detail analysis of ZBP demonstrats the non-local relation of Majorana corner mode on side surface.

**Methods**

Pyrene, NaIO$_4$, Bu$_4$NBr, Na$_2$S$_2$O$_4$, and Dodecyl bromide were purchased from Sigma Aldrich and used as received. CDCl$_3$ was purchased from Cambridge Isotope Laboratories and was used for the $^1$H NMR spectroscopic studies. A mixture of pyrene-4, 5, 9, 10-tetraone(10mmol), Bu$_4$NBr (13mmol), and Na$_2$S$_2$O$_4$ (115mmol) in H$_2$O (50ml) and THF (50ml) was shaken for 5min. Then the color of mixture was changed from dark brown to pale yellow. Bromododecane (60mmol) was added, and followed by aqueous KOH (306mmol, in 50ml H$_2$O). The mixture was stirred for overnight, poured into a mixture solution of H$_2$O (50ml) and ethyl acetate (30ml). The yellow solid was filtered and washed with ethanol. After drying in a vacuum, the solid was recrystallized from ethyl acetate resulting in a white solid with a yield of 85%. To create pelletized orthorhombic phase crystal, the orthorhombic phase crystals was placed in the evaculable pellet press for 13mm pellets (PIKE technologies) and then applied press with crusher digital hydraulic press (PIKE technologies)

Theta/2theta, out-of-plane and in-plane x-ray diffraction patterns of all samples were measured on a SMARTLAB (Rigaku Co. Ltd.,) diffractometer using monochromatized Cu-K$\alpha$ ($\lambda$ = 0.15418 nm) radiation under 40 kV and 100 mA. The dc magnetic susceptibility was measured using a SQUID-VSM (Quantum Design). The electric characterization response measurements were carried out using a semiconductor characterization system(Keithley-4200-SCS), which was connected to a probe station (MSTECH).

**Acknowledgement**

This research was supported by Basic Science Research Program through the National Research Foundation of Korea(NRF) funded by the Ministry of Education (2018R1D1A1A02047853).

**Author contributions**

Dong Hack Suh designed, initiated and directed this research. Kyoung Hwan Choi performed the experiment and analyzed the data. Dong Hack Suh and Kyoung Hwan Choi co-wrote the manuscript. All authors discussed the results.

**Competing interest.** The authors declare no competing interests.

**Supplementary information** is available for this paper